\documentclass[pra,11pt,onecolumn, showkeys,showpacs]{revtex4}

\usepackage{times}

\begin{document}

\title{Comment on "Controlled mutual quantum entity authentication using
entanglement swapping"}

\author{Gan Gao\footnote{Corresponding author. E-mail:
gaogan0556@163.com}}

\affiliation{Department of Electrical Engineering, Tongling
University, Tongling 244000, China}

\date{\today}

\begin{abstract}
Kang {\it et al.} [Chin. Phys. B {\bf 24} (2015) 090306] proposed a controlled mutual
quantum entity authentication protocol. We find that the proposed protocol is not secure,
that is, Charlie can eavesdrop the shared keys between Alice and Bob without being detected.
\end{abstract}



\maketitle

Recently, using GHZ-like states and entanglement swapping, Kang {\it et al.} proposed
a controlled mutual quantum entity authentication (CMQEA) protocol [1], in which there are three participants: Charlie, Alice and Bob. Both Alice and Bob are authentication parties, and Charlie acts as a center, similar to a telephone company, a server, etc. Although he can control all of the mutual
authentication phases, Charlie is forbidden to obtain any secret message of authentication parties. Reviewing previous papers [2-9], we see that the center, the telephone company or the server, etc. in them are similarly forbidden in that aspect. In Kang {\it et al.}'s CMQEA protocol [1], we also see that they claimed that the proposed CMQEA protocol was secure after two possible attacks, the internal attack and the external attack, were discussed. However, this is not a fact. In the Comment, we will show that the center, Charlie may eavesdrop the keys of an authentication party without
introducing any error, that is, we will prove that Kang {\it et al.}'s CMQEA protocol is not secure. Next, we start to describe Charlie's eavesdropping strategy in detail, and when describing, please refer to the phases of implementing Kang {\it et al.}'s CMQEA protocol (or Section $4$ in the paper [1]) for the clarity.

In the P2 phase, before Charlie sends the sequences including decoy qubits to Alice and Bob, he performs $\sigma_{z}$ basis measurements on qubit $C_{2i-1}$ and qubit $C_{2i}$, and Bell basis measurements on qubits $A_{2i-1}$, $A_{2i}$ and qubits $B_{2i-1}$, $B_{2i}$ (here the subscript 2i-1 and 2i designate the order of GHZ-like state in the sequence also). Since relying on decoy qubits to check the security, this action that Charlie performs these measurements ahead of schedule will not be detected by Alice and Bob in the S1 and S2 phases. Therefore, when the E1 phase goes on, if Charlie chooses Alice, obviously, she doesn't apply the Pauli operator on GHZ-like states, but Bell states now. But, Alice doesn't know this at all. When the E3 phase starts, as soon as Alice announces her measurement outcome $a_{2i-1}a_{2i}$, Charlie will easily obtain Alice's Pauli operator, that is, her keys. Later on, Charlie is required to reveal the measurement outcome of classical bit $c_{2i-1}c_{2i}$, and he may directly announce the previously obtaining outcomes. At this stage, we can't help asking that, would this introduce a error? The answer is no. The reason is as follows. We see that, when Charlie's eavesdropping exists, entanglement swapping of Bell states is finished by Charlie; when no eavesdropping exists, that is, in Kang {\it et al.}'s CMQEA protocol, it is finished by Alice and Bob. Interestingly, the measuring and deducing outcomes of Alice and Bob in the two cases are same. Therefore, they cannot detect Charlie's eavesdropping.

In conclusion, we have proposed an eavesdropping strategy on Kang {\it et al.}'s CMQEA protocol, and shown that Charlie can eavesdrop the keys of an authentication party without introducing any error. In other words, we have proved that Kang {\it et al.}'s CMQEA protocol is not secure.
\\

\noindent {\bf Acknowledgements}\\

I thank my parents for their encouragements. This work is supported
by the National Natural Science Foundation of China under grant No.
11205115, the Program for Academic Leader Reserve
Candidates in Tongling University under grant No. 2014tlxyxs30 and
the 2014-year Program for Excellent Youth Talents in University of Anhui Province. \\

\noindent {\bf References}

\noindent[1] Kang M S, Hong C H, Heo J, Lim J I and Yang H J 2015 {\it Chin. Phys. B} {\bf 24} 090306

\noindent[2] Li C Y, Zhou H Y, Wang Y and Deng F G 2005 {\it Chin.
Phys. Lett.} {\bf 22} 1049

\noindent[3] Li C Y, Li X H, Deng F G, Zhou P, Liang Y J and Zhou H Y 2006 {\it Chin.
Phys. Lett.} {\bf 23} 2896

\noindent[4] Wang W Y, Wang C, Wen K and Long G L 2007 {\it Chin.
Phys. Lett.} {\bf 24} 1463

\noindent[5] Wen X J, Liu Y and Zhou N R 2007 {\it Opt. Commun.}
{\bf 275} 278

\noindent[6] Guo Y, Shi R and Zeng G 2010 {\it Phys. Scr.} {\bf 81}
045006

\noindent[7] Gao G 2012 {\it Chin. Phys. Lett.} {\bf 29} 110305

\noindent[8] Chou Y H, Zeng G J, Lin F J, Chen C Y and Chao H
C 2014 {\it Mobile Networks and Applications} {\bf 19} 121

\noindent[9] Huang W, Wen Q Y, Liu B and Gao F 2015 {\it Chin. Phys. B} {\bf 24} 070308

\enddocument